\documentclass[preprint2]{aastex701}

\begin{document}

\title{Deuterated water ice on the satellites of Saturn}

\author[0000-0002-8255-0545]{Michael E. Brown}
 \affiliation{Division of Geological and Planetary Sciences, California Institute of Technology, Pasadena, CA 91125, USA}
 \email[show]{mbrown@caltech.edu}

\author[0000-0002-0767-8901]{Samantha K. Trumbo}
\affiliation{Department of Astronomy \& Astrophysics, University of California, San Diego, La Jolla, CA 92093, USA}
\email{satrumbo@ucsd.edu}

\author[0000-0002-7451-4704]{M. Ryleigh Davis}
\affiliation{Division of Geological and Planetary Sciences, California Institute of Technology, Pasadena, CA 91125, USA}
\email{rdavis@caltech.edu}

 \author[0000-0002-4960-3043]{Swaroop Chandra}
\affiliation{Division of Geological and Planetary Sciences, California Institute of Technology, Pasadena, CA 91125, USA}
\email{sc0296@caltech.edu}


\begin{abstract}
The deuterium to hydrogen ratio in water ice in a planetary 
body carries important information on
the history of water processing and delivery in the protostellar
nebula.
For a giant planet satellite,
the D/H ratio is also affected by the processes and temperatures of
the circumplanetary or circumstellar environment in which the satellites formed.
Here we present robust JWST spectroscopic detections of the 4.14 $\mu$m O-D stretch
absorption line (analogous to the 3 $\mu$m water O-H stretch) on the mid-sized 
Saturnian satellites and use these detections
to infer a D/H ratio on  each satellite. Within the limitations
of the technique, we find that all of the satellites are consistent
with having a D/H ratio of about $1.5 \times$ Vienna Standard Mean Ocean
Water (VSMOW), which
is about an order of magnitude higher than the value of the atmosphere 
of Saturn. 
A much higher previously reported D/H ratio for Phoebe is
ruled out at the 10$\sigma$ level, and a 3$\sigma$ upper limit
of 2.3 $\times$ VSMOW is obtained. 
The elevated D/H ratios
demonstrate that the solid planetesimals and pebbles that built the
satellites never sublimed and re-equilibrated with the gaseous
circumplanetary disk. The similarity of the D/H measurements 
across all satellites suggest that the D/H ratio of water ice in the
vicinity of Saturn at the time of satellite formation was also
approximately 1.5 $\times$ VSMOW.

\end{abstract}

\keywords{}
\section{Introduction}
Deuterated water is a powerful tracer
of the processing of interstellar ice in planetary
systems, providing a window into how interstellar ices, organics, and dust are incorporated into the disks
\citep{cleeves_ancient_2014,
yang_dh_2013, albertsson_chemodynamical_2014}.
In our own protoplanetary nebula, 
dust grains
delivered from cold molecular clouds could have
carried water ice with a D/H ratio enriched by 
orders of magnitude
above the bulk solar system value \added{of
about $2.1\times 10^{-5}$ \citep{geiss_abundances_1998}.} 
In warmer regions of the disk,
sublimation of these ices into the gas
phase would cause quick equilibration of
the D/H ratio with the bulk H$_2$ \citep{lecluse_hydrogen_1994}, leading to
water with solar D/H values.
In the outer regions of the disk,
direct incorporation of this ice into growing
bodies 
or sublimation of the ice in regions too cold
to re-equilibrate with H$_2$,
would preserve
the elevated D/H values \citep{yang_dh_2013}.
The values of D/H across the solar system thus tell
the story of transport, sublimation, and temperature
in the disk.

In the inner solar system, all water ice grains would have
completely sublimated, so D/H values should be expected to be solar.
Nonetheless, the inner solar system is enriched by about a factor of $\sim$7
compared to the solar value, \added{with a value of about 1.5$\times 10^{-4}$}, as
measured on Earth, Mars, Vesta, and C- and S- type asteroids
\citep{hallis_dh_2017}.
Such elevated values already show that
some material from interstellar ices is
eventually transported as solids into the terrestrial region,
though the precise mechanisms are debated \citep{alexander_origin_2017}.
In the colder outer portions of the disk, interstellar ices should be able to be 
more directly incorporated into icy bodies like comets, Kuiper belt obejcts,
or icy satellites.
Models for
outer solar system water ice D/H ratios predict
values
ranging from a factor of several enriched over
the terrestrial value \citep{yang_dh_2013, furuya_water_2017} to a factor of 100 over terrestrial \citep{albertsson_chemodynamical_2014} depending on
differences in \added{stellar} outflow, ice transport and incorporation, and disk temperatures.

Comets are the best studied messengers from the outer solar system,
and they have
been found to have D/H ranging from the terrestrial 
value to enrichments by about a factor of 4 
\citep[i.e.,][]{bockelee-morvan_cometary_2015, lis_terrestrial_2019, muller_high_2022}.
The sublimation and jetting processes active
on these rapidly heating comets could lead to fractionation effects that could change D/H measured in
the gaseous coma \citep{brown_experimental_2012}, making interpretation more difficult,
but it appears plausible that comets may have formed over a variety of distances and temperatures
in the nebula and that some increase in D/H with distance is present.

The D/H ratios of the satellites of the giant planets should hold additional information
on the D/H values in the middle-solar system and
on the processing of ices within the circumplanetary environments. Based on our understanding of isotopic exchange in circumstellar enviornments
\citep{yang_dh_2013, albertsson_chemodynamical_2014},
satellite formation in
a hot circumplanetary environment should lead to complete re-equilibration of
the D/H ratio to the value of the dominant H$_2$ gas, \added{which, for Jupiter and
Saturn is approximately the solar value \citep{pierel_dh_2017}}.  Temperature gradients
across the formation region could be revealed by systematic D/H gradients.
Little is known of these D/H ratios, but
using spectra from the Visual and Infrared Mapping Spectrograph (VIMS) onboard the
Cassini spacecraft, \citet[hereafter C19]{clark_isotopic_2019} demonstrated that the fundamental O-D stretch
absorption feature -- analogous to the 3 $\mu$m O-H stretch feature in water ice -- is
detectable in the rings and icy satellites of Saturn at approximately 4.14 $\mu$m. While the absorption features were near the limit
of detection for the VIMS data, new JWST observations have robustly shown the 4.14 $\mu$m feature
in the rings of Saturn \citep{hedman_waterice_2024}. Here we present JWST
reflectance spectra of the 4.14 $\mu$m region of
the icy Saturnian satellites at higher signal-to-noise and
spectral resolution than obtained for the VIMS observations.
We then discuss the implications of these detections
of deuterated water on the surfaces of these satellites
for both the formation of the solar system and the
Saturnian system.

\section{Observations and analysis}
JWST observations of the Saturnian satellites were obtained between 16-Oct-2023 and 25-Jul-2024.
The observations and data reduction are fully described in \citet{brown_jwst_2025}, so are only briefly summarized
here. Spectra of the leading and trailing hemispheres of the inner medium-sized satellites of Saturn -- Mimas,
Tethys, Dione, and Rhea -- and of the outer co-rotating satellite -- Iapetus -- as well as single
observations of the non-corotating Hyperion and Phoebe were taken using the G235H and G395M
grisms, covering the wavelength range from 1.7 to 5.2 $\mu$m, with a gap between 2.38 and 2.48 $\mu$m 
in the G235H setting. Critically, the G395M grating  does not
have the 4.08-4.26 $\mu$m gap that the higher resolution G395H grating does. The O-D 4.14 $\mu$m line 
would fall within the gap for higher resolution data. \added{The full spectrum of each satellite can
be seen in \citet{brown_jwst_2025}.}

Fig. 1 shows the JWST spectrum of the leading 
hemisphere of Dione, which is dominated by
the usual 2, 3 and 4.5 $\mu$m absorption features
of water ice. The inset shows two small absorption
features between 4.1 and 4.3 $\mu$m. 
These features are the 4.14 $\mu$m O-D stretch
absorption and the 4.26 $\mu$m CO$_2$
absorption discussed in \citet{brown_jwst_2025}.
\begin{figure}
\begin{center}
\hspace*{-1cm}\includegraphics[scale=.5]{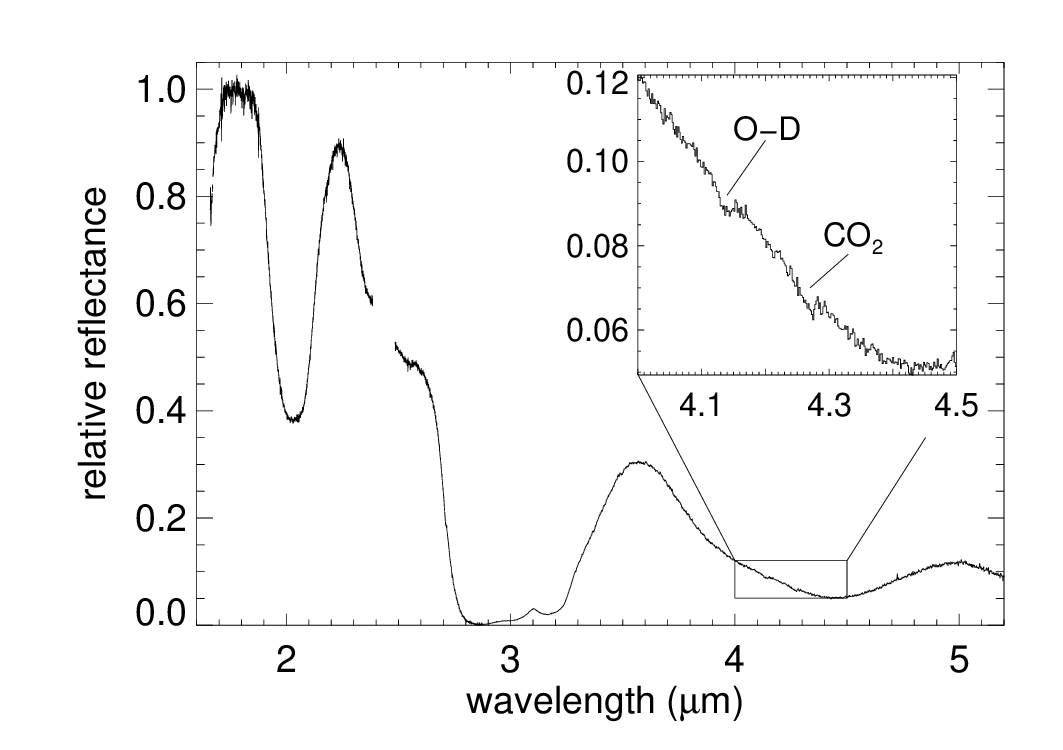}
    \caption{JWST spectrum of the leading hemisphere
    of Dione with an inset highlighting
    the 4.14 $\mu$m O-D stretch absorption
    and the 4.26 $\mu$m CO$_2$ absorption.}
    \end{center}
\end{figure}

The absorption due to the O-D stretch is seen
on nearly all of the satellites. To better visualize
each of the regions around the O-D stretch, we 
divide each spectrum by a continuum, which we 
construct by fitting the spectrum from 4.0 to 4.2 $\mu$m to
a second order polynomial while excluding the region from 4.10 to
4.17 $\mu$m. Each of the continuum-divided spectra is shown
in Fig. 2. In addition, we show a least-squares 
gaussian fit to the continuum-divided
spectrum, where we fix the width of the gaussian to be
0.0124 $\mu$m -- a value found from first allowing this parameter to 
be free and then taking the average of the results. For our least-squares
fit, we derive uncertainties in the original spectrum by calculating the
root-mean-square deviation from the spread after our continuum division.
The fractional absorption, which we define as the depth of the
absorption compared to the continuum-divided spectrum, using our
fixed gaussian width, is shown in Table 1.

\begin{deluxetable*}{lcCcCC}
\tablecaption{Measured and derived spectral parameters}
\tablehead{\colhead{satellite} 
& \colhead{longitude} &\colhead{O-D} & \colhead{2 $\mu$m} & \colhead{{\rm 4.14-to-2} $\mu$m} & \colhead{D/H} \\
& & \colhead{absorption} & \colhead{absorption} & \colhead{ratio}  & \\
  & \colhead{(deg)} & \colhead{(\%)} &\colhead{(\%)} & & \colhead{($\times$ VSMOW)}   }

\startdata
Mimas (leading)      &  52  & 4.5\pm 0.3 &  67 &  0.066\pm 0.005 &  1.6\pm 0.1\\
Mimas (trailing)     &  262 & 3.6\pm 0.3 &  60 &  0.060\pm 0.005 &  1.4\pm 0.1\\
Tethys (leading)     &  78  & 3.9\pm 0.3 &  66 &  0.059\pm 0.005 &  1.4\pm 0.1\\
Tethys (trailing)    &  271 & 3.2\pm 0.3 &  63 &  0.051\pm 0.004 &  1.2\pm 0.1\\
Dione (leading)      &  98  & 5.1\pm 0.4 &  60 &  0.085\pm 0.006 &  2.1\pm 0.2\\
Dione (trailing)     &  274 & 3.1\pm 0.3 &  45 &  0.068\pm 0.007 &  1.6\pm 0.2\\
Rhea (leading)       &  76  & 4.0\pm 0.2 &  65 &  0.061\pm 0.004 &  1.4\pm 0.1\\
Rhea (trailing)      &  261 & 3.2\pm 0.3 &  56 &  0.058\pm 0.006 &  1.3\pm 0.1\\
Hyperion             & -    & 3.1\pm0 .5 &  54 &  0.057\pm 0.009 &  1.3\pm 0.2\\
Iapetus (leading)    & 80   & 0.8\pm 0.2 &  18 &  0.045\pm 0.009 &  1.0\pm 0.2\\
Iapetus (trailing)   & 267  & 2.5\pm 0.3 &  69 &  0.036\pm 0.005 &  0.8\pm 0.1\\
Phoebe\tablenotemark{a}               & 65   & 1.3\pm 0.8 &  15 &  -  &  1.7\pm 1.1\\
\enddata
\tablecomments{
Longitude is the sub-observer longitude at the time
of observation. Hyperion has no defined longitude
system. The D/H ratio is given relative to VSMOW. }
\tablenotetext{a}{The D/H value 
for Phoebe is derived by comparison to radiative transfer models
of C19, rather than the ratio method used for the other satellites.}
\end{deluxetable*}
    
\begin{figure*}
    \begin{center}
        \includegraphics{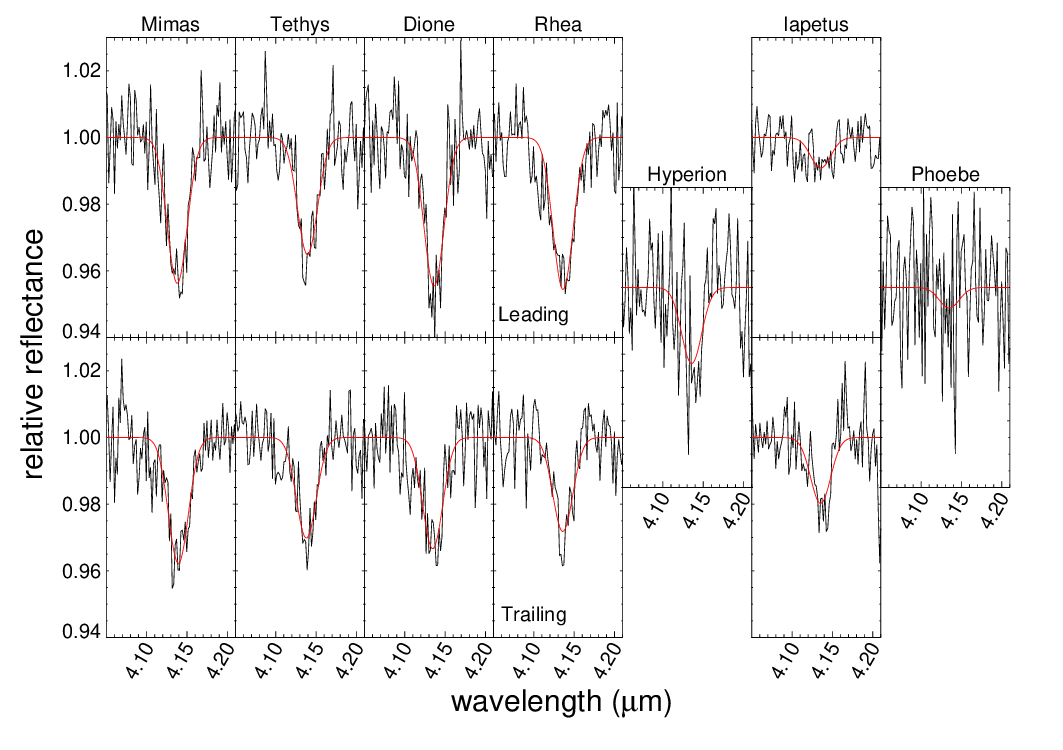}
        \caption{The continuum-divided spectra of the Saturnian satellites, 
        in the region of the O-D absorption. The red line shows a gaussian
        fit to the data using a fixed width. Most satellites have separate
        measurements for the leading and trailing hemispheres, but Hyperion and
        Phoebe, which are not synchronously rotating, only have single measurements. The O-D absorption is robustly
        detected at nearly every satellite.}
    \end{center}
\end{figure*}

\section{The D/H ratio}
While the measurements of the 4.14 $\mu$m absorption feature confirms the detection of deuterated 
water on these objects, converting this detection into a D/H ratio is more complicated. 
The detected
O-D stretch feature is analogous to the 3 $\mu$m O-H stretch for non-deuterated water. The O-H stretch
feature is saturated in all of the spectra, so a simple ratio of the 4.14 to the 3 $\mu$m depth will
not yield meaningful results. 

C19 use a reflectance spectrum
radiative transfer model to understand the spectral effects of incorporation
of deuterated water into a spectrum. For objects with significant
non-water ice components, they use this method for their final
derived D/H values. For clean water ice, they
showed from their modeling
that for grain sizes between about 5 and 100 $\mu$ -- which brackets
the range of grain sizes on these satellites --
the ratio of the 4.14 $\mu$m HDO absorption
to that of the 2 $\mu$m H$_2$O combination band stays approximately constant and can be used to
estimate the D/H ratio. 
They calibrated their 
modeling approach with three laboratory spectra of water ice with D/H values of
1, 3.2, and 16 times that of Vienna Standard Mean Ocean Water (VSMOW)  --which has
a D/H value of $1.56 \times 10^{-4}$ \citep{hagemann_absolute_1970} -- to which they fit a third-order
polynomial and suggest an accurarcy of $\sim$10\% for clean water ice. \added{As will be shown below,
the D/H ratios derived for the inner icy satellites using this method match that found
from the {\it in situ} measurements of the Enceladus plume within the uncertainties \citep{waite_jr_liquid_2009}, increasing
confidence in the technique.}

This band ratio approach has important benefits
and important limitations. The most important benefit
is that the method is straightforward, directly related to the 
observations, and easily reproducible. In addition,
the method is insensitive to
grain size over the range relevant for the Saturnian satellites, and
the ratio of the depth of the 
2 to 4.13 $\mu$m absorption lines is preserved if the water ice is linearly mixed
with a spectrally neutral material, which is possibly relevant on the
darker satellites. This band ratio is {\it not} preserved
if water ice is intimately-mixed with a spectrally neutrally material.
In this case, the band ratio can either rise, if the material is darker than the water
ice at all wavelengths, or it can fall, if the material is brighter than
the water ice. Phoebe, which is optically dark and has more muted
water signatures than the other mid-sized satellites \citep{clark_compositional_2005}, likely is effected by these issues. Other potential uncertainties can arise owing to the
different possible states of the water ice (crystalline vs. amorphous vs.
a mixture) and their possibly
different effects on
both the 2 $\mu$m and 4.14 $\mu$m features, which could affect all measurements.

With the caveats noted above in mind, we adopt the C19 ratio method
for converting the spectra to values of D/H for this initial
analysis for all satellites except for Phoebe (discussed below). 
We calculate the depth of the 2 $\mu$m absorption by fitting a 
linear continuum to the median of the spectrum between
1.79 and 1.871 $\mu$m and the median between 
2.23 and 2.26 $\mu$m, dividing the full spectrum by this continuum,
and measuring the maximum fractional depth in the region
between 1.79 and 2.24 $\mu$m. The depth of the 4.14 $\mu$m line is
taken from the gaussian fits in Figure 2. Table 1 lists these values 
for each of the satellites as well as the uncertainties for the 4.14 $\mu$m
depth. No uncertainty
is given for the 2 $\mu$m depth as our ratio uncertainties are 
dominated by the 4.14 $\mu$m depth uncertainty. 
Note that the band area ratio would make a better observational measurement,
as this ratio is unaffected by the spectral resolution of the measurements,
but here we use band depth to remain consistent with C19. The bands here are fully resolved, so the 
area and depth ratios will be the same. The D/H values derived 
from these ratios are also given in Table 1, while Figure 3 shows
both the ratio and the derived D/H for each satellite.

\begin{figure}
    \begin{center}
        \hspace*{-1 cm}\includegraphics[scale=.55]{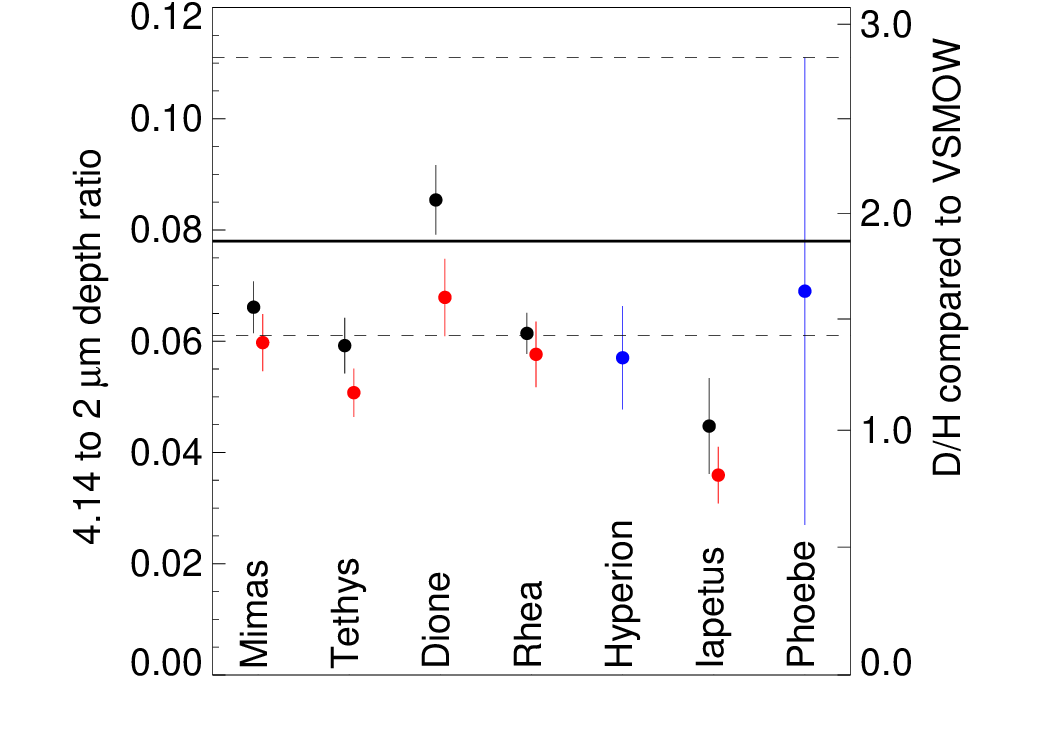}
        \caption{The ratio of the depth of the 4.14 $\mu$m
        absorption to 2 $\mu$m absorption for each
        of the satellites. The leading hemispheres are
        shown as black points while the trailing hemispheres are red. Hyperion and Phoebe,
        which are non-synchronously rotating, are
        shown in blue. The D/H value derived
        using the C19 calibration is shown
        on the right. The value measured for Enceladus
        is shown as the bold horizontal line, while the
        1$\sigma$ upper and lower limits are shown as 
        thinner dashed lines. The D/H value shown for Phoebe is
        derived from calibration to a radiative transfer model, 
        rather than from the 4.14 to 2$\mu$m depth ratio.}
    \end{center}
\end{figure}

\subsection{The inner satellites}
The D/H values derived for the inner icy satellites are consistent with the value derived in C19 for Rhea, (1.2$\pm0.2 \times$ VSMOW )
and within the uncertainties of the Cassini Ion and Neutral Mass Spectrometer (INMS) measurement
of D/H in the plume of Enceladus, which found a value of
1.85$^{+.95}_{-.45}$ compared to VSMOW \citep{waite_jr_liquid_2009}. For most satellites,
we found consistent values for the leading and trailing hemispheres.

Two exceptions to this general trend are seen. Dione has a derived
D/H value elevated above the other inner satellites, and its leading
hemisphere has a significantly higher value than its trailing
hemisphere. The implied $\sim$30\% variation between the leading and
trailing hemispheres of Dione is difficult to explain with a native
source for deuterated water. The leading hemisphere of Dione receives
significantly more mass influx from the E-ring than the trailing hemisphere\citep{kempf_saturns_2018}, which could
plausibly bring enhanced amounts of deuterated water, but we would then
expect Mimas, which receives a higher E-ring flux on the trailing hemisphere,
to have the opposite asymmetry, which is not observed.
The most conservative interpretation is that 
all of the D/H values of these inner icy satellites
are identical and that the spread seen in derived 
D/H values reflects the limitations of converting the 4.14 to 2 $\mu$m 
absorption ratio to a D/H value, \added{though it should be noted that in almost
all cases the leading and trailing hemisphere measurements are identical within the uncertainties}.  
In this interpretation,
the deuterated water detected could either be native to each
satellite or could still be brought in from the E-ring if all surfaces
are at least covered enough to be optically thick, as suggested
by radar measurements \citep{le_gall_dust_2019}.

\subsection{Hyperion and Iapetus}
Hyperion, which receives little Enceladus material but does
contain patches of what is likely Phoebe-ring dark material, has a 
derived D/H ratio consistent with those of the inner satellites. 
If the dark material on Hyperion is primarily a coating and thus spatially
segregated, the 4.14 to 2 $\mu$m ratio should accurately estimate D/H.
If, on the other hand, the surface ice of Hyperion is more intimately 
mixed with Phoebe-like dark material, this material will suppress
the 2 $\mu$m feature (where Hyperion is bright) more than the 4.14 $\mu$m
feature (where Hyperion is dark), leading to an elevated 4.14 to 2 $\mu$m 
ratio and a higher inferred value of D/H. We thus take the 
derived value for Hyperion to be an upper limit to the true value. 
More detailed modeling will be required to accurately determine
the value of D/H for Hyperion.

The derived values for the leading and trailing hemispheres of Iapetus 
are consistent in spite of the nearly factor-of-four difference
in the 2 $\mu$m band depth and factor of 2 difference in reflectance
around 4.5 $\mu$m seen by 
these observations \citep{brown_jwst_2025}. 
This consistency suggests that the band ratio method is adequately
accounting for these differences, though, like for Hyperion, we
consider the measurement for the dark leading hemisphere of
Iapetus to be a upper limit. Formally, the values derived for
Iapetus are $\sim$40\% lower than those on the inner icy satellites,
but we again conservatively interpret this spread as being with the range
of uncertainty of our approximation. As with Hyperion, detailed modeling
of the spectrum of Iapetus could help to obtain more accurate results.

\subsection{Phoebe}
C19 suggest an elevated value of D/H of $8\pm2 \times$ VSMOW for
Phoebe using VIMS data and a radiative transfer spectral model.
The details of the model are difficult to reproduce for comparison,
so
we instead chose to use a simple spectral comparison to evaluate the
claim of elevated D/H on Phoebe.
In Figure 4 we show the JWST spectrum of Phoebe, the VIMS disk-average
spectrum of Phoebe, and the averaged spectra of the other satellites.
As can be seen from the Figure, the C19 VIMS spectrum appears to have
a strong $\sim$4.15 $\mu$m absorption with an average depth of 4.5\% from
4.124 to 4.158 $\mu$m --
slightly longer wavelengths
than the D/H features seen on the other satellites by JWST. The JWST 
Phoebe spectrum rules out such a line at about the 10$\sigma$ level.

While we cannot reproduce the model of C19, if we take their modeling
as a calibration point and assume the $\sim$6\% depth of the feature
seen by VIMS indeed corresponds to a D/H ratio of $8 \times$ VSMOW,
we would infer that the $1.3\pm0.8$\% value measured for the 
4.14 $\mu$m depth at Phoebe corresponds to a value of $1.7\pm1.1 \times$
VSMOW, or, more appropriately, a 1$\sigma$ upper limit of $2.8\times$  VSMOW.
Interestingly, this value is not dissimilar to the value of $2.1^{+1.6}_{-1.4}$
that would
be derived from using the simple ratio method, lending support to our
use of that method for Hyperion and Iapetus.

\begin{figure}
    \begin{center}
        \hspace*{-1 cm}\includegraphics[scale=.5]{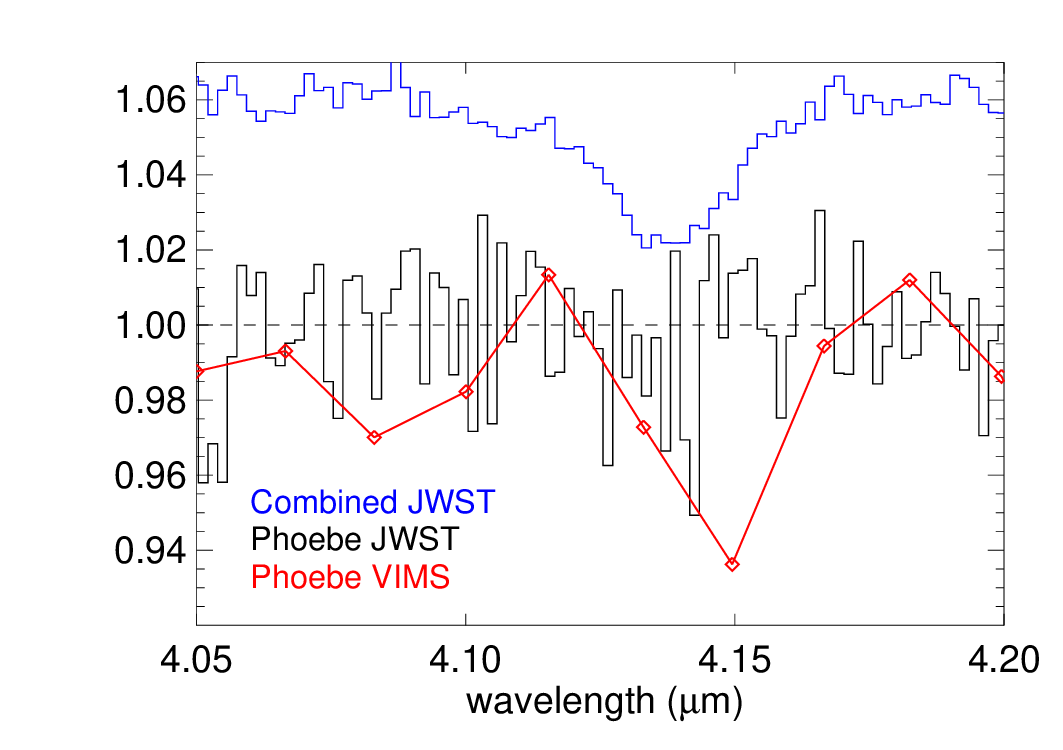}
        \caption{A comparison of the continuum-divided JWST spectrum of
        Phoebe and the VIMS spectrum of Phoebe from C19. Also shown is
        the average spectrum of all of the Saturnian satellites with 
        robust 4.14 $\mu$m absorption detections.The JWST data rule
        out an absorption with the position and depth of that seen
        by VIMS at the 10 $\sigma$ level.}
    \end{center}
\end{figure}

\section{The origins of the satellites}

The D/H ratios of the satellites of Saturn are enhanced by at least an order 
of magnitude compared to the atmosphere of Saturn, which has a D/H similar to the solar
value \citep{blake_refining_2021}. 
The strong implication of this simple fact -- even with the uncertainties
in the D/H calibration -- is that the satellites did not form
out of material that condensed out of a hot planetary sub-nebula, 
as initially explored by \citet{pollack_origin_1984}. In such a case the deuterated
water would have re-equilibrated with the much more abundant H$_2$ and the D/H value would
be quickly diluted to the Saturnian value. Multiple other lines of evidence have shown that 
\added{satellite formation in a} hot
circumplanetary nebula is implausible, however, and no modern 
formation model posits such a formation pathway.

Current models for the formation of the 
Saturnian satellites suggest either slow formation in
a gas-starved disk \citep[i.e.][]{canup_common_2006, batygin_formation_2020} or formation from reaccreated ring material, where the ring can be 
the result of a larger satellite disruption \citep{canup_origin_2010}  or the result of the tidal disruption of
a large passing icy body \citep{dones_recent_1991, hyodo_ring_2017}. All of these scenarios can be made consistent with the
approximately constant D/H value across the Saturnian system under the assumption that the planetesimals
and pebbles which form the satellites all came from regions with elevated D/H. 

In the hybrid scenario suggested by \citet{blanc_understanding_2025}
and informed by recent ideas of faster migration for the Saturnian 
satellites, large satellites form at 
some distance from Saturn and migrate inward where some are perhaps lost.
The penultimate satellite is tidally disrupted and forms the rings -- which eventually 
spawns the icy mid-sized satellites -- while the
proto-Titan survives inward migration and, upon dissipation of the disk,
tidally migrates outward. Hyperion and Iapetus presumably
also form out of this initial disk, though no satisfactory explanation
of their formation pathway has been presented \citep{castillo-rogez_origin_2018}.
In this scenario, the original satellites would have to have formed from
icy planetesimals and pebbles which did not equilibrate with the gas disk.
The lack of a rise in the D/H ratio beyond Titan
would then suggest that no strong temperature gradient existed
across the satellite formation region and that the D/H value of
the satellites largely reflects that of the planetesimals and pebbles
entering the system at the distance of Saturn.
The similarity of the D/H ratio of the icy mid-sized satellites
is a natural consequence of their origin from a single body.

The inferred upper limit for the D/H ratio of Phoebe is consistent with 
that derived for the other satellites. Phoebe appears to have a D/H 
ratio broadly consistent with the range seen in comets,
suggesting a
pre-capture origin in the outer solar system. 

\section{conclusions}
Deuterated water is detected on all of the observed mid-sized icy satellites
and on Hyperion and Iapetus. The D/H ratio on these satellites is approximately
an order of magnitude higher than in the atmosphere of Saturn, demonstrating
that the ices incorporated into these satellites never sublimated and equilibrated
with the gaseous circumplanetary nebula.
Only upper limits can be placed on Phoebe, but it cannot have a significantly
higher D/H ratio than the other satellites in the system.

The similarity in D/H ratio of all of the satellites suggests that this ratio
is representative of the D/H ratio
of the planetesimals and pebbles in the vicinity of Saturn at the time of
satellite formation. The inferred value of $\sim 1.5\times$ VSMOW is
enriched compared to the inner solar system, but does not appear to be
as high as some of the cometary values seen.

This moderate enhancement in D/H at the distance of Saturn suggests perhaps a mild
increase in D/H as a function of heliocentric distance in the solar system, 
similar to that seen in the laminar models of \citep{albertsson_chemodynamical_2014} that
do not have transport, which seems unrealistic.
Models tracking D/H ratios incorporating modern ideas of disk structure, pebble transport and accretion,
and satellite formation have yet to be constructed, but will be critical for interpreting D/H ratios
measured in planetary environments. 

One interesting prediction from these values of D/H in the ices in the vicinity of the forming
Saturn is that the ices at Uranus should be similar. Indeed, the presence of the disk
gap at Jupiter suggests that most solid material would be flowing into the Saturnian system
from beyond Saturn. Uranus, which makes no such gap, would then likely be bathed in the same
solid material. The D/H ratio at Uranus is depleted compared to VSMOW by a factor of
3.5 \citep{feuchtgruber_dh_2013}, a value from than 5 times lower than that we have inferred for
the Saturnian satellites. Uranus is expected -- though not confirmed -- to have fully mixed
its atmosphere and interior in its past, 
and thus the D/H ratio is expected to reflect the bulk average of the nebular H$_2$
and the ices that initially formed the planet. Using this expectation, \citet{feuchtgruber_dh_2013}
derive a D/H ratio for the ices that formed Uranus of a value of $\sim$0.4$\times$VSMOW, a value
more than 3.5 times lower than that we derive for the ices that formed the Saturnian satellites.
\citet{feuchtgruber_dh_2013} themselves are uncomfortable with this conclusion and suggest multiple
possible solutions, including the possibility that the interior is not mixed and that the interior
is significantly more rocky than current assumed. The high values of D/H for the Saturnian satellites
highlight this continued discrepancy.

Using the 4.14 $\mu$m O-D stretch feature to infer a D/H ratio via reflectance
spectroscopy is a powerful technique,
particularly with the advent of JWST coverage in this wavelength region. 
We caution, however, that significant laboratory and modeling work is still
needed to understand the robustness of this technique and its application
across the solar system.
While the nearly-pure water ice surfaces of the inner mid-sized satellites of
Saturn present perhaps the ideal test case for this technique, understanding how
to reliably apply the technique to more complicated surfaces remains work in progress.

\begin{acknowledgments}
We thank Francis Nimmo for conversations that led to this work.
This work is based on observations made with the NASA/ESA/CSA James Webb Space Telescope. The data were obtained from the Mikulski Archive for Space Telescopes (MAST) at the Space Telescope Science Institute, which is operated by the Association of Universities for Research in Astronomy, Inc., under NASA contract NAS 5-03127 for JWST. These observations are associated with program \#3716. Support for program \#3716 was provided by NASA through a grant from the Space Telescope Science Institute, which is operated by the Association of Universities for Research in Astronomy, Inc., under NASA contract NAS 5-03127. The specific observations analyzed can be accessed via \dataset[ DOI:10.17909/30px-vq56 ]{http:doi.org/10.17909/30px-vq56}
\end{acknowledgments}

\facilities{JWST/NIRSpec}

\bibliography{allrefs}
\bibliographystyle{aasjournal}

\end{document}